\documentclass[12pt]{article}
\pdfoutput=1
\usepackage{amsmath,amssymb}
\usepackage{graphicx}% Include figure files
\usepackage{color}
\usepackage{cases}
\newlength{\extraspace}
\newlength{\extraspaces}
\def\bsklength{.8mm} %{2mm} % for more than double spacing

\newcommand{\beq}{\begin{equation}}
\newcommand{\eeq}{\end{equation}}
\newcommand{\bseq}{\addtocounter{subeqno}{1}\begin{subequations}}
\newcommand{\eseq}{\end{subequations}}

\newcommand{\appsection}[1]{
\vspace{5mm}
\pagebreak[3]
\addtocounter{section}{1}
\setcounter{equation}{0}
\setcounter{subsection}{0}
\setcounter{figure}{0}
\phantomsection%
\addcontentsline{toc}{subsection}{\protect\numberline{\thesection.\ \ }{#1}}
\noindent{\large \bf \thesection. #1}
\nopagebreak
\medskip
\nopagebreak}

\newcommand{\BZ}{{\mathbb Z}}

\font\mathscript=eusm10 at 12pt
\font\mathscripts=eusm7
\font\mathscriptss=eusm5
\newfam\mathscri
\textfont\mathscri=\mathscript
\scriptfont\mathscri=\mathscripts
\scriptscriptfont\mathscri=\mathscriptss
\def\mathscr#1{{\fam\mathscri\relax#1}}

\font\mathfrakt=eufm10 at 12pt
\font\mathfrakts=eufm7
\font\mathfraktss=eufm5
\newfam\mathfraki
\textfont\mathfraki=\mathfrakt
\scriptfont\mathfraki=\mathfrakts
\scriptscriptfont\mathfraki=\mathfraktss
\def\mathfrak#1{{\fam\mathfraki\relax#1}}

\def\CL{{\cal L}}

\def\CO{{\cal O}}

\def\g{{\sf g}}
\def\a{{\sf a}}

\def\Det{{\rm Det}}

\def\half{{\tfrac{1}{2}}}
\def\pa{\partial}

\newcommand{\e}{{\rm e}}

\newcommand{\Tr}{{\rm Tr}}

\def\ln{{\rm ln}}

\begin{document}
\setcounter{page}{0}
\addtolength{\baselineskip}{\bsklength}
\thispagestyle{empty}
\renewcommand{\thefootnote}{\fnsymbol{footnote}}        %for symbols

\begin{flushright}
arXiv:yymm.nnnn [hep-th]\\
%{\sc DRAFT: \today}\\
\end{flushright}
\vspace{.4cm}

\begin{center}
{\Large
{\bf{Hybrid Gauge Theory}}}\\[1.2cm]
%{\large{\it{---??---}}}\\[1.2cm] %title
{\rm HoSeong La\footnote{hsla.avt@gmail.com}
}%          %author
\\[3mm]
{\it Department of Physics and Astronomy,\\[1mm]
Vanderbilt University,\\[1mm]              %address
Nashville, TN 37235, USA} \\[1.5cm]

\vfill
%{\sc Abstract}\\[1cm]
{\parbox{15cm}{
\addtolength{\baselineskip}{\bsklength}
\noindent

Cyclic symmetry $C_N$ is gauged in such a way that the local parametrization
is provided by a Lie group: matter fields are in irreducible representations of
$C_N$ while gauge fields are in the adjoint representation of a Lie group,
hence ``hybrid". Allowed simple Lie groups are only SO(2) for $N=2$, 
SU(3) for $N=3$, and SU(2) for all $N$.
The implication of the local discrete symmetry $C_N$ is evident as
the ratio of the coupling constant to the usual gauge theory one of 
the parametrization Lie group is given by that of the length between 
any two vertices of a regular N-polygon to 
the radius of the circumcircle: $2\sin(n\pi/N),\ n\in {\mathbb Z}_N$.

}
}

%{Submitted to {\it somewhere}}

\end{center}
\noindent
\vfill

%%%% table of contents %%%%
%\newpage
%\setcounter{page}{1}
%\pagenumbering{roman}
%\tableofcontents
%\vfill
%%%%%%%%%%%%%%%%%%%%%%%%%%%%%%%%%%%%%%%%

\newpage
\setcounter{page}{1}
\setcounter{section}{0}
\setcounter{equation}{0}
\setcounter{footnote}{0}
\renewcommand{\thefootnote}{\arabic{footnote}}  %for numbers
\newcounter{subeqno}
\setcounter{subeqno}{0}
\setlength{\parskip}{2mm}
\addtolength{\baselineskip}{\bsklength}

\pagenumbering{arabic}

%%%%%%%%%%%%%%%%%%%%%%%

%\newsection{Introduction}
\noindent
\underline{Introduction}

The role of continuous symmetries in Nature is very evident as the final
loophole is closed with the discovery of Higgs.
On the contrary, the role of discrete symmetries still remains as an 
elusive mystery. 
While many argue for the phenomenological roles of global discrete symmetries 
in the low energy world, there is a conflicting argument that no global 
discrete symmetry is allowed in Nature due to the quantum gravity 
effects\cite{Froggatt:1991ft}\cite{Banks:2010zn}.\footnote{One way out of 
this paradox is to treat the low energy discrete symmetries as 
emergent\cite{La:2013qia}.}

Discrete symmetries usually appear as global symmetries, 
that is, the transformations, being constant, do not depend on the spacetime. 
Nevertheless, we often use the terminologies like ``local discrete symmetry"
or ``discrete gauge symmetry".
What it means is that the actual discrete transformations are still constant 
and the same as in the global cases, but the distinction rises if we claim 
that only gauge invariant objects under this discrete symmetry are 
physically observable as in the case of continuous gauge 
symmetry\cite{Krauss:1988zc}. 
In most cases, these discrete gauge symmetries are inherited from  
continuous gauge symmetries after spontaneous symmetry
breaking. So, this type of models needs to be embedded into a 
continuum theory to justify its local nature of discrete symmetry, and
any gauge fields appearing are those of the continuous gauge symmetry,
albeit there is an extra constraint on coupling constants due to the 
discreteness of the remaining symmetry\cite{Krauss:1988zc}.

In this paper, we will propose another way of incorporating local discrete
symmetries: matter fields belong to irreducible representations of a discrete
group, while gauge fields are associated with a Lie group. Hence, we call it
``Hybrid Gauge Theory". For this, we will introduce a new mathematical 
object, which we call ``Group Family", to parametrize 
representations of a discrete group, turning into what can
be treated to be local. In short, a group family is a parametrized 
family of a discrete group, and we will define in detail momentarily.
Although this group family is not strictly a group,
it suffices to define a legitimate gauge theory.

For specific examples, we will only consider the cyclic group, $C_N$, cases. 
However, it can be generalized for other discrete groups.

\noindent
\underline{Group Family}

The basic idea stems from the fact that representations of a discrete group
are not unique.
For a given representation of a discrete group 
$G=\{\a_0,\a_1, \cdots, \a_n\}$, $\a_0=1$, 
there exist representations $G=\{\g_0,\g_1, \cdots, \g_n\}$ 
such that they are related by a similarity transformation as
\beq
\label{e1}
\g_i= U^{-1}\a_i U, \ i=0, 1,\cdots, n,
\eeq
which the identity element also trivially satisfies.
Then we can parametrize the representations in terms of parameters of $U$ as
\beq
\label{e2}
\g_i(\theta_1,\cdots, \theta_\alpha)
= U^{-1}(\theta_1,\cdots, \theta_\alpha)\a_i U(\theta_1,\cdots, \theta_\alpha).
\eeq

Note that $G$ is a group for fixed $\theta$, but not a group for 
varying $\theta$ because,
for $\g_1(U_1)$ and $\g_2(U_2)$,
\beq
\g_1(U_1)\g_2(U_2)=U_1^{-1}\a_1 U_1 U_2^{-1}\a_2 U_2 \notin G.
\eeq 
So we will use a notation $G[K(N_F)]$ to identify the object 
\beq
G[K(N_F)]=\left\langle \g_i | \g_i=U^{-1}\a_i U, \a_i\in G, U\in K(N_F)\right\rangle
\eeq
and  call it a ``group family", i.e. a family of discrete group
$G$ parametrized by a Lie group $K(N_F)$, where $N_F$ is the number of copies
of matter fields customarily called ``flavors".
$\g_i$'s with different $\theta$'s are related as
\beq
\g'_i=U'^{-1}\g_i U', \mbox{ for } U'\in K(N_F).
\eeq

To obtain a local symmetry, all we need is to assign different gauge parameters
at different spacetime points. 
Note that local group elements at different spacetime points
do not have to be related by the same group transformations.
Therefore, as long as $\theta(x)$ varies continuously over the spacetime,
we have differentiable local discrete transformations. 
Then we can define a gauge theory with a local discrete symmetry
in terms of these parametrizations. 
%The number of gauge fields should be
%the same as that of independent parameters. 
Ideally, the number of flavors $N_F$ could be identified as that of 
irreducible representations, however, other cases are also interesting so that
we will consider as well. $N_F$ also controls the parametrization Lie group.

Note that $\g^\dagger\g=1$ implies that $U^\dagger U=1$ and $\a^\dagger\a=1$
so that the parametrization group $K(N_F)$ is necessarily an orthogonal or
a unitary group. 

\noindent
\underline{Warm-up: $C_2[{\rm SO}(2)]$}

Consider the simplest discrete group $C_2=\{1,\a\}$, 
a cyclic group of order two,
which has just one nontrivial group element, $\a$, in addition to the identity. 
There are two irreducible representations, so it is natural to take 
two flavor case ($N_F=2$) and a two dimensional representation of $\a$ will 
be sufficient such that
\beq
\label{e3a}
\a=\sigma_3 =
\left(
\begin{array}{cc}
1 & 0\\ 
0 & -1
\end{array}
\right), 
\eeq
where we will use the convenient Pauli matrix notation from now on. 
In the $C_2$ case, $\a^\dagger=\a$, which is a useful property for $\a$
to be a symmetry generating transformation. The similarity
transformation $U$ in this case should just have one parameter to 
sustain the abelian nature of $C_2$ and the corresponding $\g$ should be real.
That naturally fixes $U$ to be the SO(2) rotational matrix such that
\beq
\label{e3b}
U(\theta)=\e^{i\sigma_2\theta}=
\left(
\begin{array}{cc}
\cos \theta & \sin \theta\\ 
-\sin \theta &\cos \theta
\end{array}
\right).
%\left(
%\begin{array}{cc}
%c & s\\ 
%-s &c
%\end{array}
%\right).
\eeq
Then\cite{La:2013rra}\cite{La:2013qia}
\beq
\label{e4}
\g=U^{-1}\a U=
\left(
\begin{array}{cc}
\cos  2\theta & \sin  2\theta\\ 
\sin  2\theta &-\cos  2\theta
\end{array}
\right).
\eeq
Note that $\g^2=1$, $\g^\dagger=\g$ and $\Det\g=-1$.
Then $C_2[{\rm SO}(2)]=\{1,\g(\theta)\}$ in our notation.

Now consider two scalar matter fields $\Phi\equiv(\phi_1,\phi_2)^T$, 
then $C_2[{\rm SO}(2)]$ acts as 
\beq
\label{e5}
\Phi'=\g\Phi.
\eeq
such that the bilinear form is invariant as
\beq
\label{e6x}
\Phi'^T\Phi'=\Phi^T\g^2 \Phi=\Phi^T\Phi.
\eeq
In this parametrization, the gauging can be achieved by demanding
\beq
\label{e6}
D'_\mu\Phi'=\g D_\mu\Phi,
\eeq
where
\beq
\label{e7}
D_\mu \equiv \pa_\mu -iQ A_\mu \sigma_2
%\quad 
%\sigma_2=
%\left(
%\begin{array}{cc}
%0 & -i\\ 
%i &0
%\end{array}
%\right),
\eeq
such that $Q=\zeta q$ with $\zeta\neq 1$ for unit charge $q$ will indicate
the nontrivial charge condition matching the discrete symmetry.
Using
\bseq
\begin{align}
\label{e11a}
\g\sigma_2\g &=-\sigma_2, \\
\label{e11b}
\g\pa_\mu\g  &= i2\pa_\mu \theta\sigma_2,
\end{align}
\eseq
we can obtain the transformation rule for the gauge field $A_\mu$ as
\beq
\label{e12}
A'_\mu =-A_\mu-\tfrac{2}{Q}\pa_\mu\theta
=-A_\mu-\tfrac{1}{q}\pa_\mu\theta,
\eeq
where the charge $Q$ now should be constrained with $\zeta=2$ as $Q=2q$.
This gauge transformation property is something new because of the minus
sign in front of $A_\mu$.
Note that eq.(\ref{e11b}) is the reason why we choose $A_\mu$ in the 
representation $\sigma_2$. Even if we choose $\a=\sigma_1$, eq.(\ref{e7}) 
and eq.(\ref{e12}) remain the same. 

Together with eq.(\ref{e12}) that shows how the gauge field transforms
under $C_2[{\rm SO}(2)]$, eq.(\ref{e6}) implies that
\beq
\label{e13}
\CL[\Phi]=-\tfrac{1}{4}F^2
+(D_\mu\Phi)^T(D_\mu\Phi)-V(\Phi^T\Phi)
\eeq
is invariant under the group family $C_2[{\rm SO}(2)]$.
%This lagrangian also has a global O(2) symmetry.
This can be easily generalized to the fermionic case with the same covariant
derivative eq.(\ref{e7}) with two fermion flavors $\Psi=(\psi_1,\psi_2)^T$.
%The lagrangian is
%\beq
%\label{e14}
%\CL[\Psi]=-\tfrac{1}{4}F^2
%+i\bar{\Psi}^T\gamma^\mu D_\mu\Psi-m\bar{\Psi}^T\Psi.
%\eeq
%Two fermion flavors must have degenerate masses.

Having constructed the lagrangian, we cannot help but notice 
that it also has the usual SO(2) gauge symmetry
\bseq
\begin{align}
\label{e:13x}
\Phi' &= \e^{i2\sigma_2\theta}\Phi, \\
\label{e:13y}
A'_\mu &=A_\mu-\tfrac{1}{q}\pa_\mu\theta. 
\end{align}
\eseq
So we need to elaborate what is going on.

\noindent
\underline{Compared with SO(2), what is the difference?}

Since the local parametrization is mainly due to SO(2), 
one may wonder if $C_2[{\rm SO}(2)]$
gauge theory is anything different from SO(2) gauge theory. 
After all, the $C_2[{\rm SO}(2)]$ invariant lagrangian has SO(2) 
gauge symmetry, too.
%Note that ${\rm O}(2)=C_2\times {\rm SO}(2)$.
To see the difference we need to analyze the 
$C_2[{\rm SO}(2)]$ gauge transformation more carefully.

Most of all, $C_2[{\rm SO}(2)]$ has a peculiar property unlike SO(2).
Compared with SO(2) in which gauge transformations satisfy
\beq
U(\theta_1) U(\theta_2) =U(\theta_2) U(\theta_1)=U(\theta_1+\theta_2) \in 
{\rm SO}(2),
\eeq
in the $C_2[{\rm SO}(2)]$ case a multiplication of two arbitrary group elements
does not result in another group element, unless two gauge parameters are
the same:
\beq
\g(\theta_1) \g(\theta_2)=U( 2\theta_1- 2\theta_2) \notin C_2[{\rm SO}(2)]
\mbox{ if } \theta_1\neq \theta_2.
\eeq
This is because there is only one $\g(\theta)$ at a given spacetime point.
Elements with different gauge parameters are related in terms of
SO(2) such that
\beq
\g(\theta_2)
=U^{-1}(\theta_2-\theta_1)\g(\theta_1) U(\theta_2-\theta_1).
\eeq
Because of this,
even though ${\rm SO}(2)$ is not a subgroup of $C_2[{\rm SO}(2)]$, 
it coincidentally allows the SO(2) gauge field 
to take the role of the gauge field as well.
So, in some sense, the local $C_2[{\rm SO}(2)]$ is a partially 
augmented group family of global $C_2$ by local SO(2). 
However, the distinction is mostly in the coupling constants of the lagrangian:
In the SO(2) gauge theory, $Q=2q$ is just a choice, while in the 
$C_2[{\rm SO}(2)]$ gauge theory, it is required.

Another way to see the main difference between
$C_2[{\rm SO}(2)]$ and SO(2) gauge transformations,
let us factorize the nontrivial element of $C_2[{\rm SO}(2)]$, 
from eqs.(\ref{e3b})(\ref{e4}), as
\beq
\label{e15}
\g(\theta)=\sigma_3\, U( 2\theta).
\eeq
This does not mean $C_2[{\rm SO}(2)]=\{1,\g(\theta)\}$ is isomorphic to 
${\rm O}(2)=C_2\times {\rm SO}(2)$ because SO(2) is not a subgroup of 
$C_2[{\rm SO}(2)]$.
Now, the anatomy of the gauge field transformation reveals that
\beq
\label{e16}
A_\mu\xrightarrow{\sigma_3} -A_\mu 
\xrightarrow{{\rm SO}(2)}-A_\mu-\tfrac{1}{q}\pa_\mu \theta .
\eeq
Under just SO(2), the gauge field 
transforms as $A'_\mu =A_\mu-\tfrac{1}{q}\pa_\mu\theta$ such that,
when the gauge parameter $\theta$ is constant, the gauge field
does not change, while in the $C_2[{\rm SO}(2)]$ case it flips sign as 
$A_\mu\to -A_\mu$.
The key difference is that $C_2$ acts nontrivially on the gauge field.
But the gauging leads to a part of O(2) gauge theory\cite{Kiskis:1978ed}
except, in addition, we have a nontrivial charge condition on matter 
fields: $Q=2q$.

One may wonder that $A_\mu\to -A_\mu$ may lead to the inconsistency of 
the sign of the charge. But we believe this is not a correct interpretation.
The correct physics is always to consider the matter couplings. 
It is a ``flavor" symmetry, so it should not mean 
anything without matter couplings. Since matter-gauge interaction is 
invariant, there is no ambiguity of the sign of the charge.
So, unlike SO(2) or U(1) gauge theories, the field
strength is no longer a gauge invariant observable (nor is the case of
non-abelian gauge fields), but this is all right because any phenomenon 
of gauge theory is only observable via matter-gauge interactions.

\noindent
\underline{Generic Structure for Cyclic Groups $C_N$}

For arbitrary cyclic groups $C_N$, we need matter fields 
transforming as
\beq
\label{e:17}
\Phi'=\g\Phi,\quad \Phi'^\dagger=\Phi^\dagger\g^\dagger,
\eeq
and the hermitian conjugate satisfies $\a^\dagger=\a^{N-1}$ so that
$\a^N=1=\a\a^\dagger$, where it is sufficient to use the 
generator of $C_N$
\beq
\label{e:18}
\a={\rm diag}(1, \omega_N,\cdots, \omega_N^{N-1}), \quad
\omega_N\equiv \e^{i{2\pi\over N}}
\mbox{ such that }
\omega_N^N=1.
\eeq
In the $C_2$ case, $\g^\dagger=\g$.
Infinitesimally, for $U=\exp(i\theta_A T_A)=1+ i\theta_A T_A+\CO(\theta^2)$,
where $T_A$'s are generators of parametrization Lie group $K(N_F)$,
$\g$ reads
\beq
\label{e:21a}
\g 
%=(1-i\theta^a T^a)\a (1+i\theta^a T^a)
=\a +i\theta_A [\a, T_A]+{\rm h.o.}.
\eeq
Now we can immediately observe the fact that diagonal generators and
off-diagonal generators behave differently because $[\a, T_i]=0$, 
while $[\a, T_a]\neq 0$, where the subscript ``$i$" labels diagonal generators 
and ``$a$" the off-diagonal ones. Particularly, note that
only off-diagonal generators show up at the leading order of $\theta$.

The gauge fields transform as
\bseq
\begin{align}
\label{e:23p}
\Sigma_a A'^a_\mu+ T^i A'^i_\mu
&=\g(\Sigma_a A_\mu^a +T^i A^i_\mu-\tfrac{i}{Q}\pa_\mu)\g^{-1},\\
\label{e:23x}
\Sigma_a F'^a_{\mu\nu}+ T_i F'^i_{\mu\nu}
&= \g \left(\Sigma_a  F^a_{\mu\nu}+ T_iF^i_{\mu\nu}\right)\g^{-1}.
\end{align}
\eseq 
To find the representations for gauge fields, generalizing eq.(\ref{e11b}),
we need
\beq
\label{e:22b}
\g\pa_\mu\g^{-1} 
%=\g[\g^{-1}, U^{-1}\pa_\mu U]
%=U^{-1}\pa_\mu U - \g U^{-1}\pa_\mu U \g^{-1}
=i\pa_\mu\theta_a \a[\a^{-1}, T_a]+{\rm h.o.}.
\eeq
Then, with comparing eq.(\ref{e:22b}) to eq.(\ref{e:23p}), 
it is natural to define
\beq
\label{e:24}
\Sigma_a\equiv
\tfrac{1}{\zeta_a}\,[\a^{-1}, T_a]\a 
= \tfrac{1}{\zeta_a}\a^{-1}[T_a,\a] \mbox{ for } [T_a,\a]\neq 0
\eeq
with suitable normalizations $1/\zeta_a$ to make $\Tr \Sigma_a^2=\Tr T_a^2$
and $\zeta_a$ takes the role of the charge condition.
This is compatible with eq.(\ref{e11b}) for $\zeta=2$
in the $C_2$ case because $U=\exp(i\theta\sigma_2)$, $\a=\sigma_3$, and
\beq
\label{e:25}
\Sigma=\tfrac{1}{2}[\sigma_3,\sigma_2]\sigma_3=-\sigma_2.
\eeq
$\Sigma_a$'s do not form a closed algebra because $[\Sigma_a,\Sigma_b]$ 
produces $T_i$, but they do with $T_i$'s added. Then the same lagrangian
given in terms of $\Phi$ and $A_\mu$ would have a usual local gauge symmetry
$K(N_F)$ generated by $\{\Sigma_a, T_i\}$, but it is different from
the $C_N[K(N_F)]$ symmetry. In other words, there are two different
local gauge symmetry for this type of lagrangians, however certain constraints
are only evident in the $C_N[K(N_F)]$ symmetry, as we will see.

With eq.(\ref{e:24}), we can now obtain
\bseq
\begin{align}
\label{e:27a}
\g &=\a\left(1-i\zeta_a\theta_a\Sigma_a\right)+{\rm h.o.},\\
\label{e:27b}
\g\Sigma_c\g^{-1}
&=\a\Sigma_c\a^{-1} 
+i\zeta_b\theta_b\a[\Sigma_c,\Sigma_b]\a^{-1}+{\rm h.o.},\\
\label{e:27c}
\g T_i \g^{-1}
&=T_i +i\zeta_b\theta_b \a [T_i,\Sigma_b]\a^{-1}+{\rm h.o.}
\end{align}
\eseq
and eq.(\ref{e:23p}) becomes, with introducing short-hand notations,
\beq
A'_\mu=\a \left(A_\mu +\tfrac{1}{Q}[D_\mu, \zeta\theta]\right)\a^{-1}
+{\rm h.o.},
\quad \zeta\theta\equiv \zeta_a\theta_a\Sigma_a.
\eeq
with
\beq
\label{e:23b}
D_\mu\equiv\pa_\mu+iQ A_\mu, \quad
A_\mu\equiv \Sigma_a A^a_\mu+T_i A^i_{\mu\nu}.
\eeq
In the leading order of $\theta$, only gauge parameters for off-diagonal 
generators appear, but the missing $\theta_i$ will appear at one higher order, 
so there are still complete gauge parameters.
Note that, to preserve the original periodicity of $\theta$'s for $U\in K(N_F)$,
$\theta$'s should be still the gauge parameters for $C_N[K(N_F)]$, hence
$Q$ should cancel $\zeta$'s. This is possible only if  
$\zeta_a$'s are the same for all generators such that
\beq
\label{e:30}
Q=\zeta g, \quad \zeta_a=\zeta\ \mbox{for all}\ a,
\eeq
where $g$ is the coupling constant of the usual gauge theory 
with $\{T_A\}$ for a Lie group $K(N_F)$. 
$\zeta$ is the ratio of coupling constant of $C_N[K(N_F)]$ 
to that of the gauge theory of the parametrization Lie group $K(N_F)$. 
So, $\zeta\neq 1$ indicates that the coupling constant is not the same as 
that of the gauge theory of $K(N_F)$. 
This is also evident in eq.(\ref{e:27a}).
This also implies that the matter 
fields carry certain charges w.r.t. a U(1) subgroup generated by 
a diagonal generator due to the discrete nature of
the symmetry, similar to the $C_2[{\rm SO}(2)]$ case. 
Next, we will compute these charge conditions explicitly
to find out a rather interesting geometrical origin.

\noindent
\underline{Geometric Charge Quantization Conditions for $C_N$}

To obtain the charge condition for $C_N$, it is sufficient to consider
$C_N[{\rm SO}(2)]$ as we could easily check the fact that off-diagonal 
generators are grouped as those of SU(2) subsets 
and they can be paired as $\sigma_2$ and $\sigma_1=-i\sigma_3\sigma_2$.
So the normalization condition of $\Sigma$ in $C_N[{\rm SO}(2)]$ will be
the same as that of $C_N[{\rm SU}(N)]$ when the conventional prefactor 
$1/2$ for ${\rm SU}(N)$ generators is properly taken into account,
which is $\Tr \Sigma_a^2=\Tr T_a^2=1/2$ for $C_N[{\rm SU}(N)]$.

For now, let
\beq
\label{e:c1}
\a=
\left(
\begin{array}{cc}
1 & 0\\ 
0 &  \omega_N^n
\end{array}
\right),\quad  n=\BZ_N\setminus{\{0\}},\quad \omega_N\equiv \e^{i2\pi/N}
\eeq
and choose group family $C_N[{\rm SO(2)}]$, then
\beq
\g\pa_\mu\g^{-1}
=i\zeta_{N,n}\pa_\mu\theta \a\Sigma\a^{-1}+{\rm h.o.},
\eeq
where
\beq
\Sigma
=\tfrac{1}{\zeta_{N,n}}
\left(\a^{-1}\sigma_2 \a-\sigma_2\right)
=\tfrac{1}{\zeta_{N,n}}
\left(
\begin{array}{cc}
0 & i(1- \omega_N^n)\\ 
-i(1- \omega_N^{n*}) & 0
\end{array}
\right).
\eeq
The charge condition $\zeta_{N,n}$ can be estimated, with
demanding the norm of $\Sigma$ to be unity so that $\Tr \Sigma^2=2$, as
\beq
\label{e:c2}
\zeta_{N,n}=|1- \omega_N^n|^{1/2} 
=2\sin{n\pi\over N}
\eeq
and that $\Sigma$ simplifies as
\beq
\Sigma=
\left(
\begin{array}{cc}
0 & \omega_N^{n/2}\\ 
\omega_N^{*n/2} & 0
\end{array}
\right).
\eeq
Note that, for $n=1$, $\zeta_{N,1}$ is nothing but the ratio of the side 
length of a regular $N$-gon (regular $N$-polygon) to the radius of the 
circumcircle.
For arbitrary $n\in\BZ_N$, $\zeta_{N,n}$ are the the ratios of any lengths 
between two vertices of a regular N-gon to the radius of the circumcircle.
So this is quite a geometrical outcome.
It is consistent with the previous examples $C_2[{\rm SO}(2)]$ since 
$\zeta_{2,1}=2$. 

In eq.(\ref{e:30}), we argued that the charge conditions should be the same
for all generators to be consistent. Knowing eq.(\ref{e:c2}) is the length
between any vertices of a regular $N$-gon, we can easily figure out which
cases lead to all identical lengths. From geometry alone, they are $N=2$ and 
$N=3$. However, as we will see, as long as the parametrization Lie group is
SU(2), any $N$ is possible because different $n$'s do not appear at the same 
time. So, for simple Lie groups, allowed ones are only SO(2) for $N=2$, 
SU(3) for $N=3$, and SU(2) for all $N$. In the former two cases, the number
of the flavors is the same as that of irreducible representations of $C_N$.
If we allow semi-simple Lie groups, as long as they are products of 
these allowed simple Lie groups, they could be allowed. 

\noindent
\underline{$C_N[{\rm SU}(2)]$}

With just two flavors, we can have fixed charge conditions even for any 
$N\geq 3$, so $C_N[{\rm SU}(2)]$ is a very informative case.
To argue for the necessity of $C_N[{\rm SU}(2)]$,
let us first clarify why $C_N[{\rm SO}(2)]$ does not lead to consistent
gauge theory if $N\neq 2$.
The gauge transformation relates $\Sigma$ and $\a\Sigma\a^{-1}$ in the
limit of vanishing $\theta$, however,
\beq
\a\Sigma\a^{-1}=\Sigma^*
\eeq
implies
\beq
A'_\mu=\omega_N^n A_\mu=\omega_N^{*n} A_\mu.
\eeq
This is possible only if $\omega_N^n=\omega_N^{*n}$, which occurs if
$N=2n$, i.e. $\omega_N^n=\omega_N^{*n}=\pm 1$. In this case, eq.(\ref{e:c1})
implies $\a\in C_2$. So if $N\neq 2$, we need two gauge fields with
two generators according to
\beq
\label{e:e0}
A_\mu^1\Sigma_1+A_\mu^2\Sigma_2,
\eeq
hence leading to off-diagonal $\Sigma$'s of SU(2).

Using eq.(\ref{e:c1}), from eq.(\ref{e:24}) we can obtain
\beq
\label{e:e1}
\Sigma_1=\half
\left(
\begin{array}{cc}
0 & i\omega_N^{n/2}\\ 
-i\omega_N^{*n/2} & 0
\end{array}
\right),\quad
\Sigma_2=\half
\left(
\begin{array}{cc}
0 & \omega_N^{n/2}\\ 
\omega_N^{*n/2} & 0
\end{array}
\right),
\eeq
which, together with $T_3$, form an su(2) algebra with the same 
structure constants as $T_A$'s. Now eq.(\ref{e:e0}) can be expressed as
\beq
\label{e:e2}
\left(
\begin{array}{cc}
0 & \omega_N^{n/2}A_\mu^-\\ 
\omega_N^{*n/2}A_\mu^+ & 0
\end{array}
\right),
\quad
A_\mu^\pm\equiv \half(A_\mu^2\mp A_\mu^1).
\eeq
Then, under $C_N[{\rm SU}(2)]$ we have a consistent gauge transformation 
property
\beq
A'^{+}_\mu =\omega_N^n A_\mu^+ +\cdots
\eeq
and its complex conjugate. They appear to be charged gauge fields w.r.t.
U(1) subgroup generated by $T_3$.

Since $\{\Sigma_a, T_3\}$ form an su(2) algebra, the $C_N[{\rm SU}(2)]$
invariant lagrangian also has SU(2) gauge symmetry. Furthermore, 
since $\{\Sigma_a, T_3\}$ leads to the same structure constants as
the usual SU(2) generators $\{T_A\}$ based on the Pauli matrices do,
$\{\Sigma_a, T_3\}$ can be related to $\{T_A\}$ by the following similarity 
transformation
\beq
V^{-1} T_a V= \Sigma_a,\quad
V=
\left(
\begin{array}{cc}
1 & 0\\ 
0 &  i\omega_N^{n/2}
\end{array}
\right),
\eeq
which $T_3$ trivially satisfies, too. 
%while
%\beq
%\Sigma_1 = -\a^{-1/2} T_2 \a^{1/2},
%\quad
%\Sigma_2 = \a^{-1/2} T_1 \a^{1/2}.
%\eeq
So this SU(2) symmetry is equivalent to the usual SU(2) gauge symmetry. 

Note that for $N\geq 3$ we end up with nonabelian parametrization.
This is quite intriguing because, after all, $C_N$ is an abelian discrete
group, but we are led to non-abelian parametrization Lie groups.
Clearly, what we have is quite different from the local discrete symmetry of
Krauss-Wilczek\cite{Krauss:1988zc}. So we will briefly compare to that next.

\noindent
\underline{Comparison to Krauss-Wilczek\cite{Krauss:1988zc}}

In \cite{Krauss:1988zc} Krauss and Wilczek introduced a clarifying concept 
of local discrete symmetry such that physical observables should be invariant
under this discrete symmetry, which works as follows.
With two flavors of different charges $\Phi=(\phi_1,\phi_2)^T$, the acting
gauge transformation is given by
\beq
\g=
\left(
\begin{array}{cc}
\e^{i\theta} & 0\\ 
0 & \e^{iN\theta}
\end{array}
\right),
\eeq
then
\beq
\g\pa_\mu\g^{-1}=-i\pa_\mu\theta\zeta\Sigma,
\eeq
where $\Sigma=\mathbf{1}$ and the charge condition is
\beq
\zeta=
\left(
\begin{array}{cc}
1 & 0\\ 
0 & N
\end{array}
\right).
\eeq
The gauge field is abelian, hence U(1), because
\bseq
\begin{align}
A'_\mu\zeta &= \g (A_\mu\zeta -i\pa_\mu)\g^{-1}
=\zeta (A_\mu-\pa_\mu\theta), \\
i.e.\qquad
A'_\mu &=A_\mu-\pa_\mu\theta.
\end{align}
\eseq
In this case, $\BZ_N$ symmetry appears as the invariance under
\beq
\theta\to \theta+\frac{2\pi n}{N} \quad {\rm for}\ n\in \BZ_N.
\eeq
When $\phi_1$ gets vev, the U(1) symmetry will be spontaneously broken,
yet $\BZ_N$ symmetry remains unbroken. This is identified as
local $\BZ_N$ symmetry and $\BZ_N\subset {\rm U}(1)$.

Another way of seeing this is to express it in terms of 
$\varphi\equiv -i\ln\phi_2$ such that 
$\pa_\mu\varphi-iN A_\mu$ becomes invariant 
under $\varphi\to \varphi +N\theta$ and $A_\mu\to A_\mu-i\pa_\mu\theta$
\cite{Banks:1989ag}.

In the $C_2[{\rm SO}(2)]$ case, as we can see from 
eqs.(\ref{e:13x})(\ref{e:13y}), it is comparable to the $\BZ_2$ case
of the Krauss-Wilczek's because the $C_2[{\rm SO}(2)]$ invariant lagrangian
happens to have SO(2) gauge symmetry with $\zeta=2$,
except that spontaneous breaking of SO(2) is not needed to show the local
nature of the discrete symmetry. 
In other cases, it is different. In our case, for nonreal
\beq
\Sigma 
=\left(
\begin{array}{cc}
0 & \beta\\ 
\beta^* & 0
\end{array}
\right),\quad
\a\Sigma\a^{-1}
=\left(
\begin{array}{cc}
0 & \omega_N^*\beta\\ 
\omega_N\beta^* & 0
\end{array}
\right),
\eeq
the gauge fields must satisfy, in the limit of vanishing gauge parameters,
\beq
\left(
\begin{array}{cc}
0 & \beta A'_\mu\\ 
\beta^* A'^*_\mu & 0
\end{array}
\right)
=
\left(
\begin{array}{cc}
0 & \omega_N^*\beta A_\mu\\ 
\omega_N\beta^* A_\mu^* & 0
\end{array}
\right).
\eeq
Then, unless $\omega_2=\omega^*_2$,
the corresponding gauge parameters need to be of the form
\beq
\Sigma\pa_\mu\theta +i\sigma_3\Sigma \pa_\mu\xi.
\eeq
Since $[\Sigma, \sigma_3]\neq 0$, third generator has to be introduced
so that there is no abelian parametrization for $N\geq 3$ in our case. 
($N=2$ is a special case because $\omega_2=\omega^*_2=-1$ so that
$A_\mu=A_\mu^*$ and abelian parametrization is allowed.)
%This is because our gauging leads to a natural
%realization of $C_N\times {\rm U}(1)^{N-1}\subset {\rm SU}(N)$.

\noindent
\underline{$C_3[{\rm SU}(3)]$}

Since SU(3) is the largest possible simple Lie group leading 
to a fixed charge condition, 
which can be easily seen from the geometry of a regular triangle,
let us check out the $C_3$ case in detail.
With $\zeta_{3,1}=\zeta_{3,2}=\sqrt{3}$, SU(3) generators $\{T_A\}$
based on Gell-Mann matrices, and eq.(\ref{e:18}) for $N=3$,
eq.(\ref{e:24}) leads to 
\beq
\begin{aligned}
\Sigma_1 =
\half \left(
\begin{array}{ccc}
0 & -i \omega_3^2 &0\\ 
i \omega_3 & 0 & 0 \\
0 & 0 &0
\end{array}
\right),\ 
&
\Sigma_2 =
\half \left(
\begin{array}{ccc}
0 & - \omega_3^2 &0\\ 
- \omega_3 & 0 & 0 \\
0 & 0 &0
\end{array}
\right),\ 
\Sigma_4 =
\half \left(
\begin{array}{ccc}
0 & 0 & i \omega_3\\ 
0 & 0 & 0 \\
-i \omega_3^2 & 0 &0
\end{array}
\right),
\\
\Sigma_5 =
\half \left(
\begin{array}{ccc}
0 & 0 &  \omega_3\\ 
0 & 0 & 0 \\
 \omega_3^2 & 0 &0
\end{array}
\right),\ 
&
\Sigma_6 =
\half \left(
\begin{array}{ccc}
0 & 0 &0\\ 
0 & 0 & -i \omega_3^2\\
0 & i \omega_3 &0
\end{array}
\right),\ 
\Sigma_7 =
\half \left(
\begin{array}{ccc}
0 & 0 & 0\\ 
0 & 0 & - \omega_3^2\\
0 & - \omega_3 &0
\end{array}
\right). %\\&\quad  \Sigma_3 = 0 =\Sigma_8. 
\end{aligned}
\eeq
Again, to form a closed algebra, we need to add $T_3$ and $T_8$ of 
diagonal SU(3) generators.
We can check if $\{\Sigma_a, T_i\}$ forms the same su(3) algebra as $T_A$'s do. 
The structure constants for $\{\Sigma_a, T_i\}$ are
\beq
g_{123}=1,\ g_{164}=g_{175}=g_{247}=g_{265}=g_{345}=g_{376}=\half,\
g_{458}=g_{678}=\tfrac{\sqrt{3}}{2},
\eeq
and all others vanish. Compared with this, the su(3) algebra based on 
Gell-Mann matrices have nonvanishing structure constants
\beq
f_{123}=1,\ 
f_{147}=f_{165}=f_{246}=f_{257}=f_{345}=f_{376}=\half,\ 
f_{458}=f_{678}=\tfrac{\sqrt{3}}{2}.
\eeq
The structure constants are slightly different, unlike 
the $C_N[{\rm SU}(2)]$ case. This indicates that there may not be a similarity
transformation preserving the same su(3) algebra between the two bases.
Indeed, we can reproduce the two sets of structure constants with the following
similarity transformation:
\bseq
\begin{align}
V^{-1}T_a V &=\Sigma_a, a=1,2,4,5,\\
V^{-1}T_6 V &=\Sigma_7,\\
\label{e:s3c}
V^{-1}T_7 V &=-\Sigma_6, 
\end{align}
\eseq
and diagonal generators are invariant,
where
\beq
V=\left(
\begin{array}{ccc}
1 & 0 & 0\\ 
0 &  - i\omega_3^2 & 0\\
0 & 0 & i \omega_3 
\end{array}
\right).
\eeq
However, due to the minus sign in eq.(\ref{e:s3c}), the two su(3) algebras
are not related by a similarity transformation.
If these were related with a single sign as
$
\Sigma_a=V^{-1} T_b V,
$
the lagrangian might be identified as having the usual color
SU(3) gauge symmetry combined with $C_3[{\rm SU}(3)]$ discrete symmetry. 
So, even though the $C_3[{\rm SU}(3)]$ invariant lagrangian has another 
SU(3) gauge symmetry
generated by $\{\Sigma_a, T_i\}$, but not the same as the usual color SU(3).

We are particularly interested in the $\theta$-independent terms of
gauge transformations in eq.(\ref{e:27b}), which read
\beq
\begin{aligned}
\a\Sigma_a\a^{-1} &= -\Sigma_a^*, && \mbox{for } a=1,4,6, \\
\a\Sigma_a\a^{-1} &= \Sigma_a^*, && \mbox{for } a=2,5,7.
\end{aligned}
\eeq
Unlike in the $C_2$ case, in the $C_3$ case an SO(3) parametrization 
is not allowed because it leads to inconsistent transformation rules 
for each components as in the $C_N[{\rm SO}(2)]$ case for $N\geq 3$. 
Therefore, in the $C_3$ case, to be consistent, the parametrization Lie
group should be SU(3). Since complex gauge fields take the role of charged
gauge fields w.r.t. the diagonal generators,
this provides natural manifestation of 
$C_3\times {\rm U}(1)^2 \subset {\rm SU}(3)$.
Each complex gauge field now picks up phases $\omega_3$ or $\omega_3^2$.
Compared with the $C_2$ case in which gauge fields flips sign, 
i.e. $\omega_2=-1$,
now we have nontrivial phases showing up upon gauge transformations.
This also does not lead to any inconsistency because it does not affect
observable S-matrix of matter-gauge interactions, which are gauge invariant. 

\noindent
\underline{Final Remarks}

We have presented a consistent gauge theory different from the usual one
based on a Lie group. The most interesting outcome is the symmetry leads to
charges quantized according to the geometry of the discrete symmetry 
we start with, i.e. charges correspond to the lengths between any
two vertices of a regular polygon. 
Although it is not clear if there is any real world physical system 
directly based on the structure presented here, but we believe this
is quite an interesting theoretical outcome.
The key implication of Krauss-Wilczek's gauged discrete symmetry is
the possibility of nontrivial charges on a black hole associated with
the discrete symmetry\cite{Preskill:1990bm}. So, we believe there could be some 
physical applications in the physics we have not encountered yet.
Also a ${\rm U}(N)$ gauge symmetry can appear as $N$ D-branes are 
bounded together. This naturally has $C_N$ discrete symmetry built in. 
So there could be some connection with the stringy world.
There are only two cases that the number of flavors can be the same as that of
irreducible representations of the cyclic group $C_N$: 
$C_2[{\rm SO}(2)]$ and $C_3[{\rm SU}(3)]$.
In particular, it is also interesting to observe that the largest simple 
Lie group SU(3) is allowed with only three flavors. 
Also it will be interesting to speculate if 
this has anything to do with QCD in a certain limit, when three flavors
in our case are treated as three colors.

Another noticeable aspect of the outcome is that
the same lagrangian has a usual local gauge symmetry $K(N_F)$ 
generated by $\{\Sigma_a, T_i\}$, which is different from the 
$C_N[K(N_F)]$ symmetry. 
In other words, there are two different local gauge symmetries for the same lagrangian. In the SU(2) cases, the two gauge symmetries are related by
a similarity transformation, but not in the SU(3) case.
In any case, certain constraints are only evident in the $C_N[K(N_F)]$ symmetry. 
This raises a possibility that known SU(2) gauge theories may have another
hidden gauge symmetry, which may better explain certain phenomena.

In this paper, we have adopted 
parametrization groups mixing all flavors at once, but, in principle, 
we can mix only some of them, which leads to, e.g. 
$C_N[{\rm SU}(2)\times {\rm SU}(2)\times \cdots]$ type of group family.
We can also generalize to discrete groups other than cyclic groups.
For example, a dihedral group of order three, $D_3$, can lead to  
$D_3[K(N_F)]$ with a suitable semi-simple Lie group, e.g. 
$D_3[{\rm SU}(2)\times {\rm SU}(3)]$.
% $D_3[{\rm SU}(2)\times {\rm SU}(2)]$
%or even $D_3[{\rm SU}(2)\times {\rm SU}(3)]$ group families.
 
It will be extremely interesting if there is a mechanism to spontaneously
break the symmetry we have introduced here to lead to useful global discrete
symmetries in Nature. This will justify the origin of the unbroken global
discrete symmetries in the low energy world despite any quantum gravity effects. 

\noindent
{\bf Acknowledgments:}
I would like to thank Tom Kephart for conversations on related issues
and reading the earlier drafts.

%\newpage
\renewcommand{\Large}{\large}

\end{document}